\documentclass[%
 reprint,
showpacs,preprintnumbers,
 amsmath,amssymb,
 aps,
]{revtex4-1}

\usepackage{graphicx}% Include figure files
\usepackage{dcolumn}% Align table columns on decimal point
\usepackage{bm}% bold math
\usepackage{multirow}
\usepackage{float}
\begin{document}

%\preprint{APS/123-QED}

\title{Nuclear Weak Rates and Detailed Balance in Stellar Conditions}% Force line breaks with \\

\author{G. Wendell Misch}
\affiliation{Department of Physics and Astronomy, Shanghai Jiao Tong University, Shanghai 200240, China}
\affiliation{Collaborative Innovation Center of IFSA (CICIFSA), Shanghai Jiao Tong University, Shanghai 200240, China}

\date{\today}

\begin{abstract}
Detailed balance is often invoked in discussions of nuclear weak transitions in astrophysical environments.  Satisfaction of detailed balance is rightly touted as a virtue of some methods of computing nuclear transition strengths, but we argue that it need not necessarily be strictly obeyed, especially when the system is far from weak equilibrium.  We present the results of shell model calculations of nuclear weak strengths in both charged current and neutral current channels at astrophysical temperatures.  Using these strengths to compute some reaction rates, we find that, despite some violation of detailed balance, our method is robust up to high temperature, and we comment on the relationship between detailed balance and weak equilibrium in astrophysical conditions.
\end{abstract}

% PACS, the Physics and Astronomy
                             % Classification Scheme.
%\keywords{Suggested keywords}%Use showkeys class option if keyword
                              %display desired
\maketitle

%\tableofcontents

\section{\label{sec:intro}Introduction}

Weak interactions play crucial roles in the evolution of stars of all masses.  Main sequence stars convert hydrogen to helium via either the p-p chain or the C-N-O cycle.  In stars with mass greater than $\sim 12~M_\odot$ (solar masses), beginning with core carbon fusion, energy and entropy are carried out of the core almost entirely by neutrinos.  Up until silicon burning, most of these neutrinos are produced as pairs via thermal processes in the plasma \cite{itoh-etal:1996,pl:2015}.  However, in late silicon burning and during supernova core collapse, nuclei play an important--and eventually dominant--role in neutrino production \cite{arnett:1977,bbal:1979,omk:2004a,langanke:2015,asakura-etal:2016}.  Stars with mass $\sim 8-12~M_\odot$ may collapse due to loss of electron pressure support as nuclei in the $A=20-24$ mass range capture nuclei \cite{nomoto:1987,huedepohl:2010,martinez-pinedo-etal:2014}.  Nuclear $\beta$-decay rates are also critical in nucleosynthesis processes, influencing waiting points in the r-process (rapid neutron capture) and rp-process (rapid proton capture), as well as in neutron star cooling and type Ia supernovae \cite{kratz-etal:1986,kratz:1988,lpph:1991,pa:1992,engel-etal:1999,ml:1999b,sag:2005,bazin-etal:2008,mori-etal:2016}.  In light of the pervasive influence of nuclear weak interactions, we naturally desire the most accurate rates possible.

\begin{figure}
\centering
\includegraphics[scale=.3]{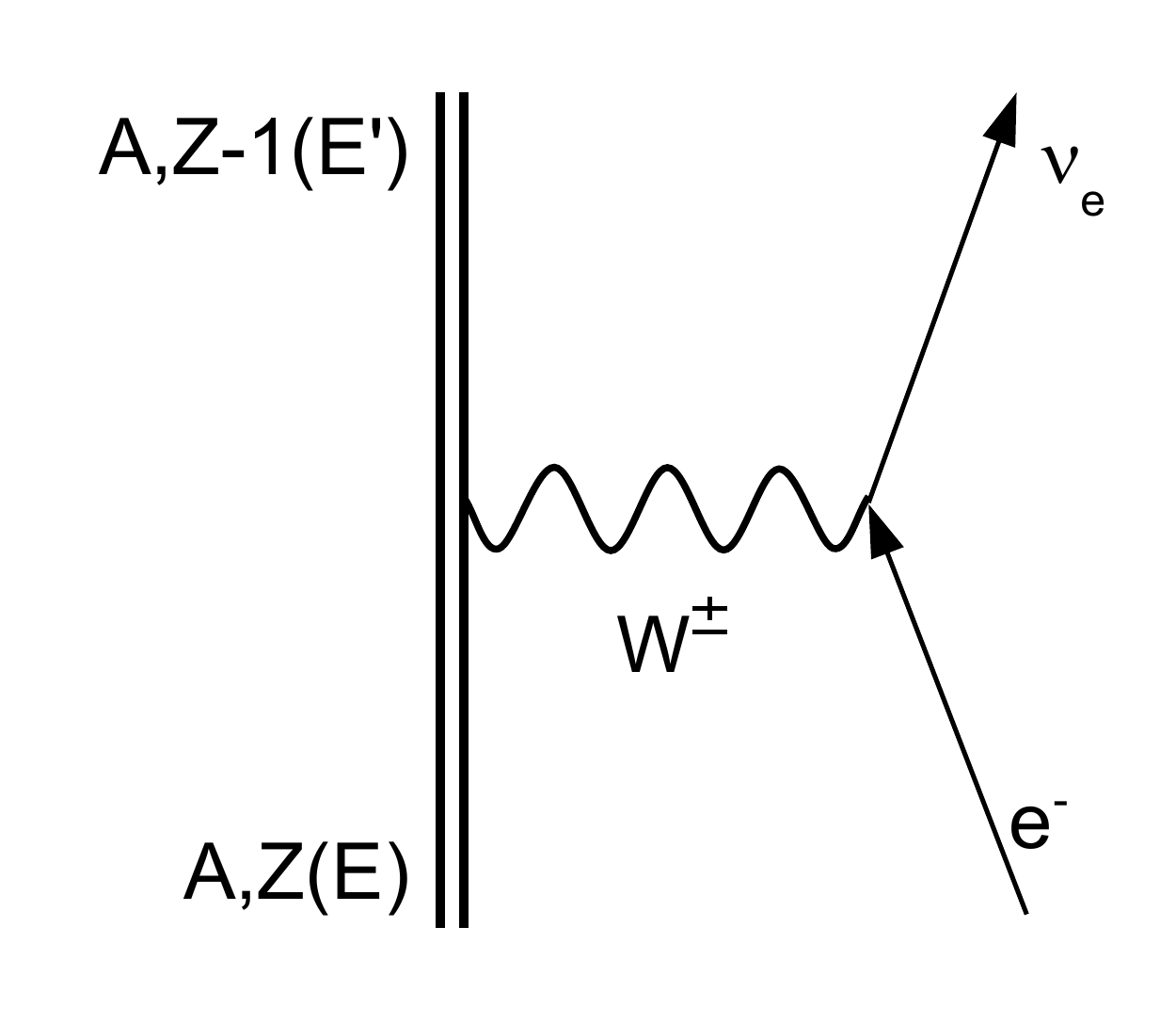}
\includegraphics[scale=.3]{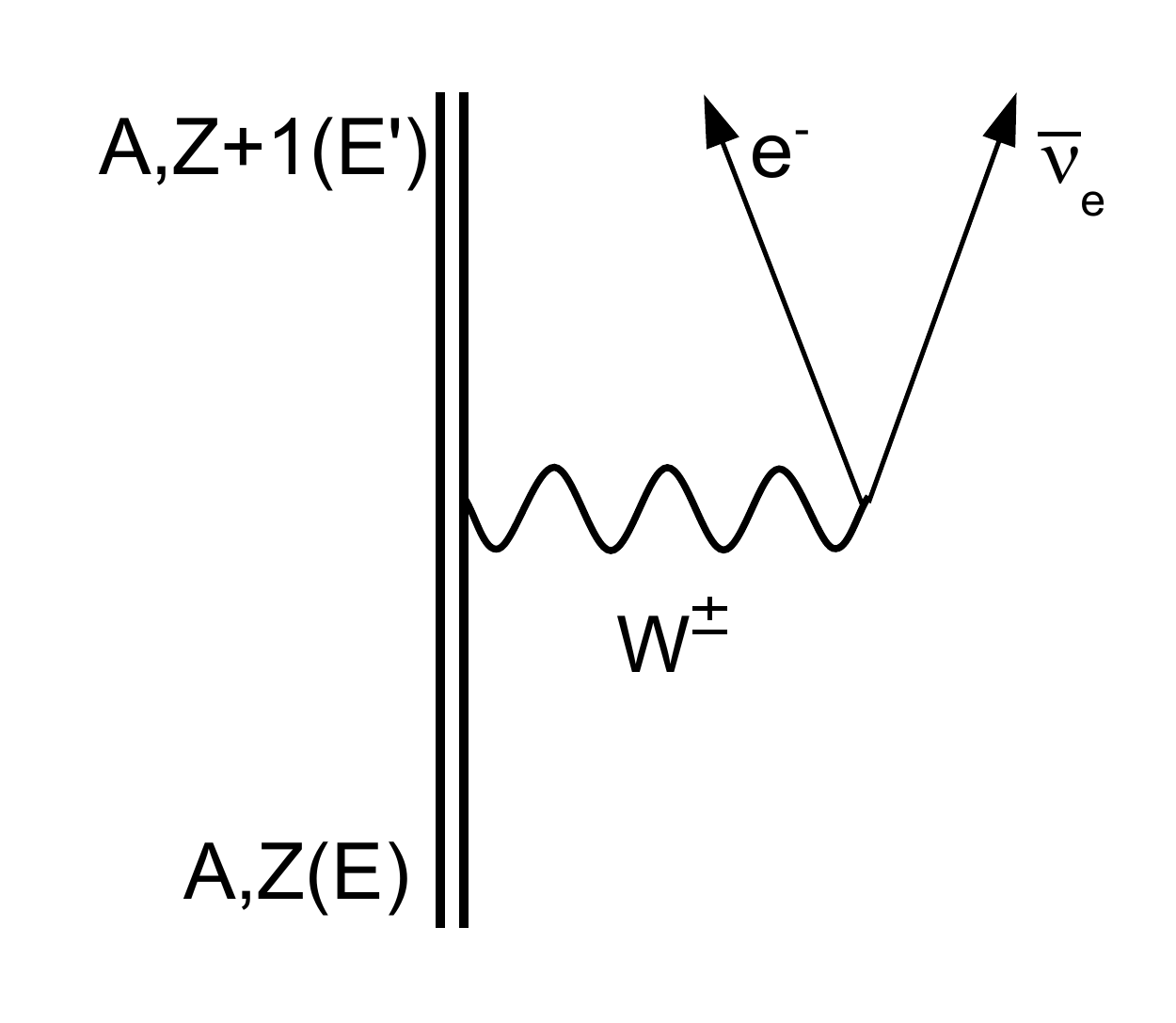}
\includegraphics[scale=.3]{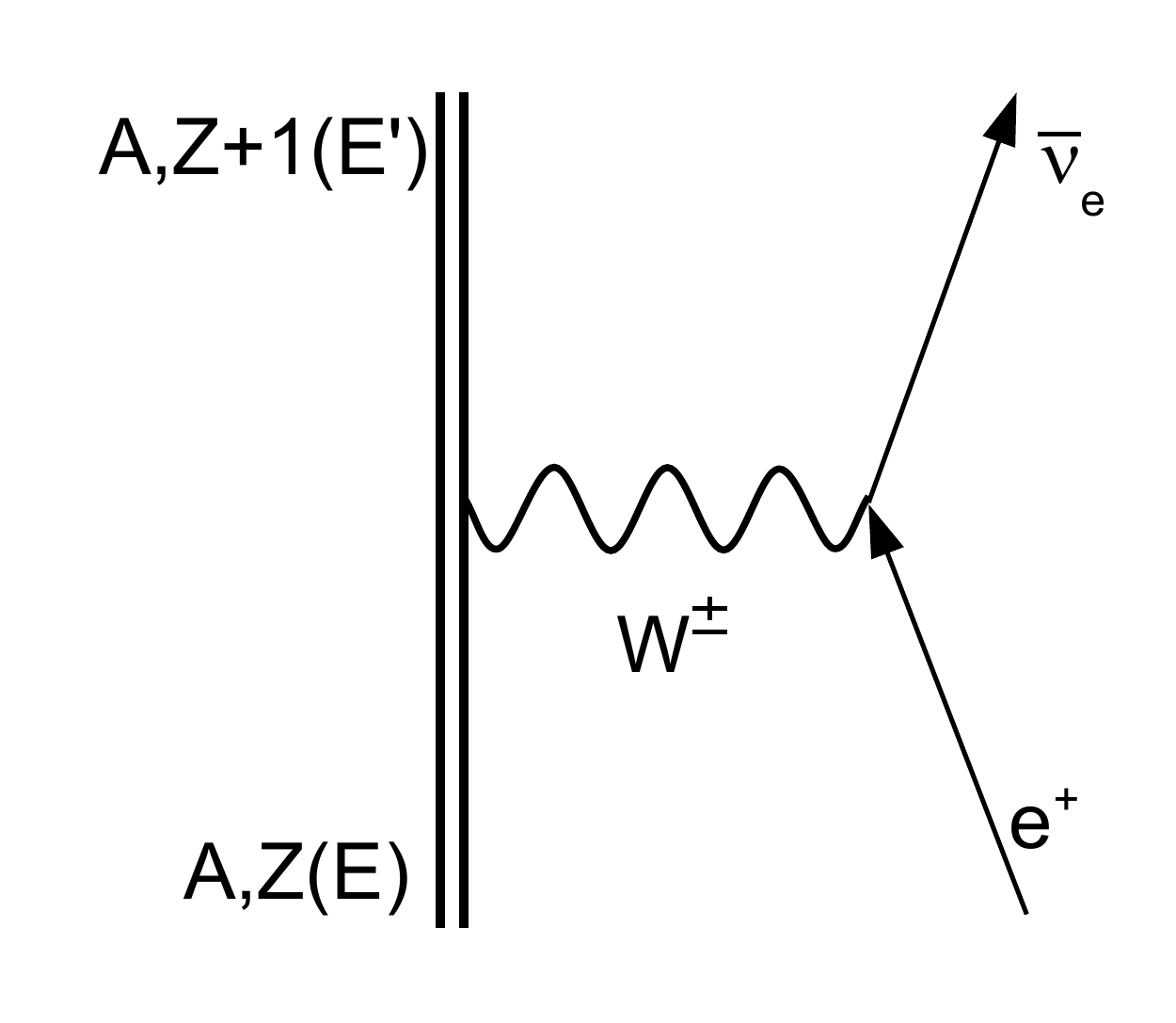}
\includegraphics[scale=.3]{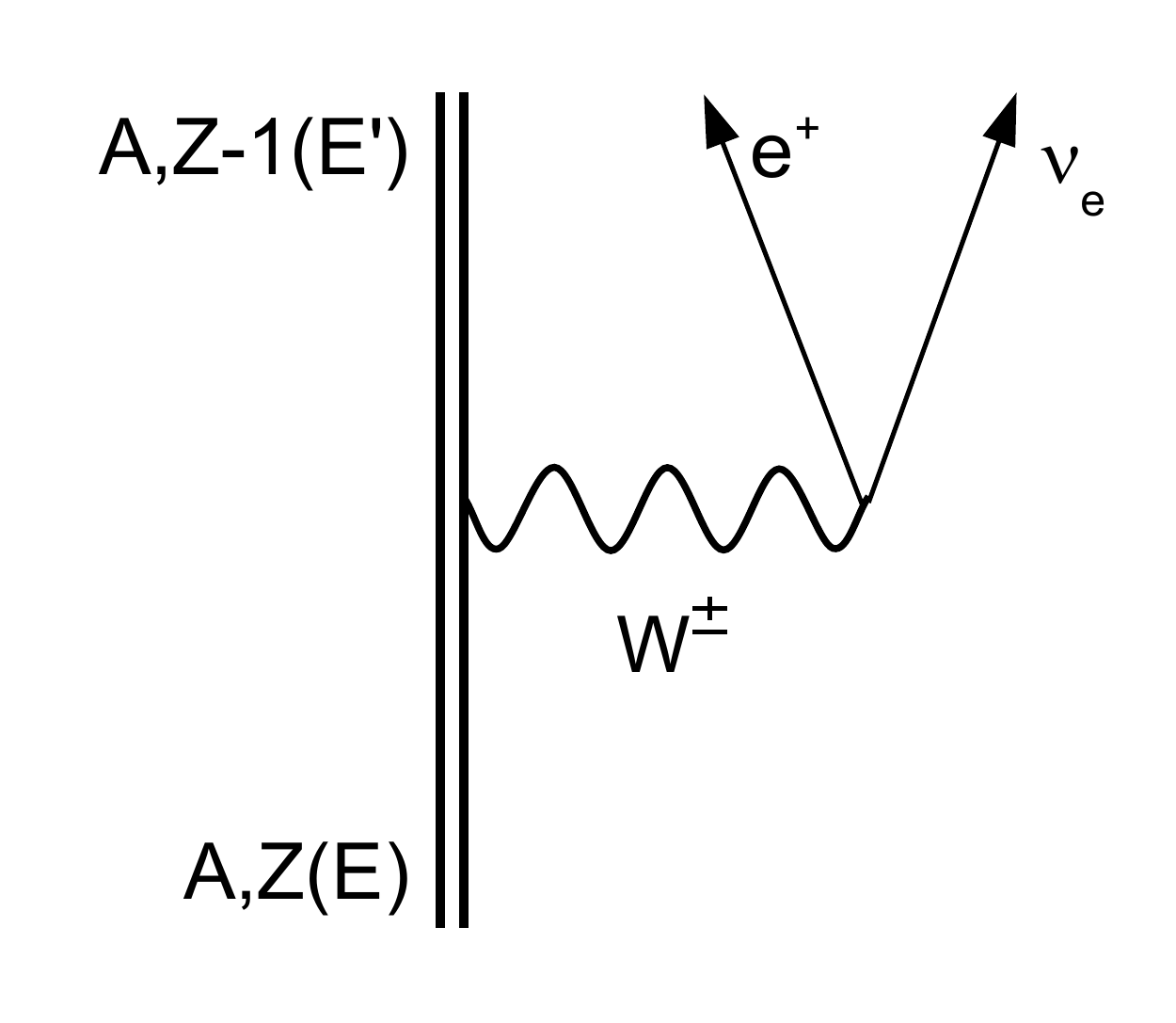}
\includegraphics[scale=.3]{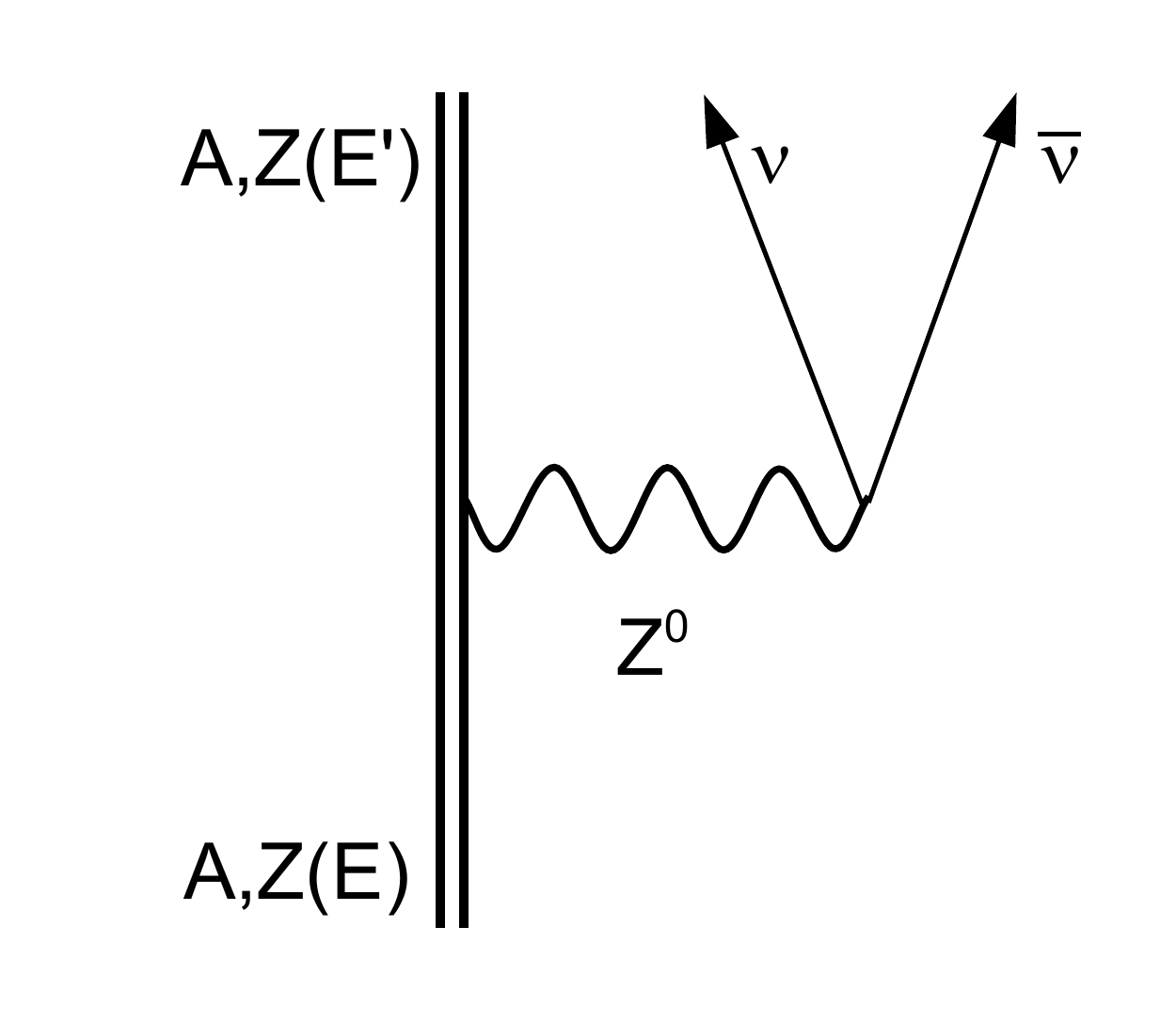}
\caption{Principle nuclear weak processes in stellar environments: electron capture, electron emission, positron capture, positron emission, and neutral current deexcitation.  Initial and final energies can both be excited states.}
\label{fig:feyn_weak_processes}
\end{figure}

There are five principle nuclear weak interaction processes in stellar environments--four charged current reactions and one neutral current: electron capture, electron emission (beta decay), positron capture, positron emission, and neutral current de-excitation, shown schematically in figure \ref{fig:feyn_weak_processes}.  The last of these is similar to nuclear gamma-ray emission, but instead of a photon, the nucleus emits a virtual Z$^0$ boson that decays into a neutrino-antineutrino pair.  Under conditions with large neutrino fluxes, the neutrinos in the Feynman diagrams of figure \ref{fig:feyn_weak_processes} can be changed from outgoing to incoming.  These processes are all sensitive to the masses and structure of nuclei.

Unfortunately, nuclear properties are notoriously difficult to accurately compute, particularly at finite temperature, and nuclear neutrino rates and energy spectra depend sensitively on these properties.  The difficulty lies particularly in the limitations of computers, as nuclei are complex many-body quantum mechanical systems that must be solved numerically.  Researchers have developed numerous techniques over the years to simplify the problem and make good approximations, gradually yielding more reliable results.

Detailed balance refers to the relationship between the forward and reverse reaction strengths of a quantum mechanical system: the forward and reverse transition amplitudes from an initial state $\vert i\rangle$ to a final state $\vert f\rangle$ are identical.  We show in section \ref{sec:strength_imbalance} how to express this for a nucleus in a thermal bath.  Because a realistic model must obey detailed balance, it is often invoked to simplify calculations, sparing the invoker direct computation of the reverse reaction strengths and ensuring that at least this aspect of their model is realistic.

Fuller, Fowler, and Newman (FFN) \cite{ffn:1980,ffn:1982a,ffn:1982b,ffn:1985} performed one of the earliest broad surveys of charged current nuclear weak rates considered accurate enough for use over the past few decades.  Their work employed the standard of using experimentally measured nuclear energies and transition strengths were available and using approximation techniques to fill the gaps.  In particular, they assumed all allowed transitions with unknown strength to have log$(ft_{ij})$ values of 5; $ft_{ij}$ is the comparative half-life of the transition from initial nuclear state $\vert i\rangle$ to final nuclear state $\vert j\rangle$ and is related to the nuclear weak interaction matrix elements \cite{bcw:1978}.  Then, they adapted the Brink-Axel hypothesis \cite{brink:thesis,axel:1962} to charged current weak interactions by assuming that the Gamow-Teller resonance occurred in excited states with the same strength and same relative transition energy as in the ground state.

This obeys detailed balance, but the downsides of the FFN technique are twofold.  First, there can be significant variation in the allowed transition strengths, so the assumption of log$(ft_{if})=5$ for all unknown transitions only holds in an average sense.  Second, the adaptation of the Brink-Axel hypothesis to weak interaction strength breaks down for initial state excitation energies of more than a few MeV \cite{mfb:2014}.

Oda et al \cite{oda-etal:1994} sought to circumvent these issues by performing large-scale shell model calculations on $sd$-shell nuclei (nuclei with mass numbers $A=17-39$).  In their calculations, they supplemented known experimental energies and strengths with computations of the 100 lowest-lying states in each $sd$-shell nucleus found using the shell model.  Because all 100 states are included as both initial and final states and no additional states are considered, this method obeys detailed balance.  It is effective over a broad range of conditions, but it does not include negative parity states, and it may miss some important strength to and from higher energy states at extreme temperatures and densities.

Caurier et all \cite{clmn:1999} and Langanke \& Martinez-Pinedo \cite{lm:2000,lm:2001} performed large scale shell model calculations on nuclei in the mass range $A=45-65$.  Unfortunately, shell model calculations quickly become intractable as nuclear mass number increases (the number of basis states grows exponentially), so this technique is restricted to lighter nuclei unless some kind of truncation is imposed.  In this case, the authors of those works considered only relatively low-lying nuclear energies of a few MeV.

In order to approach heavier nuclei, some authors have employed the quasiparticle random-phase approximation \cite{dzhioev-etal:2010,sarriguren:2013,dvw:2015}.  While this approach is good at capturing the distribution of the bulk of the nuclear transition strength and obeys detailed balance, it is not as good as the shell model at finding detailed nuclear structure (to which weak rates are sensitive).  Others have used Monte Carlo shell model methods \cite{jklo:1992,kdl:1997}.

The difficulties of heavy nuclei necessitate the development of modern techniques to treat nuclear many body problems efficiently.  Considering that most nuclei in the nuclear chart are deformed, working with a deformed basis instead of a spherical basis is more computationally economical.  However, angular momentum is not a good quantum number in a deformed basis and needs to be restored by using the angular momentum projection method \cite{sun:2016}.  The projected shell model does this \cite{hs:1995} and has been successful in describing many structural properties of deformed nuclei, making it a promising approach.  It has not yet been broadly employed to compute charged current weak interactions, but recent developments indicate that it might be very useful for just that in the near future \cite{gsc:2006,wsg:2017}.

The conclusion we must ultimately arrive at is that as of now, we simply cannot precisely calculate nuclear weak rates (and nuclear neutrino spectra) under all conditions, so we must make decisions about what to sacrifice.  This paper will make the case that among the things we can sacrifice is strict adherence to detailed balance of thermal nuclear weak strength, particularly when the environment of interest is out of weak equilibrium.

In section \ref{sec:strength_imbalance} we define nuclear weak interaction thermal strengths and and the quantity imbalance.  Section \ref{sec:calculations} describes a method of producing thermal strengths from shell model calculations and shows the results for several nuclei along with the associated imbalances.  Section \ref{sec:phase_space} defines phase space factors, and in section \ref{sec:rates} we present rate calculations for $^{32}$P using two different but related methods.  In section \ref{sec:discussion}, we talk about environments far from weak equilibrium and how that relates to detailed balance.

\section{Thermal Strength and Imbalance}
\label{sec:strength_imbalance}

In the language of reference \cite{flm:2013}, detailed balance of the neutral current thermal strength is expressed as

\begin{equation}
S^{GT3}\left(T,\Delta E\right)=e^{\Delta E/T}S^{GT3}\left(T,-\Delta E\right)
\label{eq:nc_balance}
\end{equation}
where $T$ is the temperature and $\Delta E$ is the nuclear transition energy (final excitation energy minus initial excitation).  $S^{GT3}$ is the neutral current thermal strength, given by

\begin{equation}
\begin{split}
&S^{GT3}\left(T,\Delta E\right)=\\
&\frac{1}{G(T)}\int_0^\infty dE(2J+1)\rho(E,J)e^{-E/T}B^{GT3}(E,\Delta E)
\end{split}
\label{eq:nc_thermal_strength}
\end{equation}
where $G(T)$ is the nuclear partition function, $E$ is the initial nuclear excitation energy, $J$ is the nuclear spin, $\rho(E,J)$ is the density of nuclear states, and $B^{GT3}(E,\Delta E)$ is the strength for a transition from a state with energy $E$ and a transition energy of $\Delta E$.

We refer to the thermal strength on the left hand side of equation \ref{eq:nc_balance} as that of the forward reaction, and the strength on the right hand side as corresponding to the reverse reaction.  In keeping with the appeal of symmetry, we put the forward and reverse reactions on equal footing by rewriting equation \ref{eq:nc_balance} as

\begin{equation}
e^{-\Delta E/2T}S^{GT3}\left(T,\Delta E\right)=e^{\Delta E/2T}S^{GT3}\left(T,-\Delta E\right).
\label{eq:nc_balance2}
\end{equation}

Equation \ref{eq:nc_balance2} has some conceptual subtleties that are worth touching on.  Naturally, it applies to all neutral current nuclear interactions, including lepton pair emission, lepton pair absorption, and scattering.  Notably, the reverse reaction thermal strength is not considered directly, but rather is mirrored about $\Delta E=0$.  We interpret the physical meaning of this as comparing ``up transitions'' (transitions with positive $\Delta E$, where the nucleus gains energy) in the forward reaction directly with the corresponding ``down transitions'' (negative $\Delta E$) in the reverse reaction, and vice versa.  

At finite temperature, the forward reaction will have non-zero strength for both up and down transitions (implying the same for the reverse reaction).  Depending on the reactions, this can be physically absurd; obviously, neutrino pair absorption cannot have negative nuclear transition energy, and pair emission cannot induce the nucleus to gain energy.  If the reactions under consideration do not allow up (or down) transitions, then simply neglect the corresponding domain of $\Delta E$ as being unphysical.  This argument also applies to charged current interactions.  For the sake of completeness, in our discussions of thermal strength, we will remain agnostic as to the specific reaction, and figures will contain the full domain of $\Delta E$.  Note, though, that in this work we always consider reverse reactions with the $-\Delta E$ argument as in equation \ref{eq:nc_balance2}; this can make confusing-looking reverse reaction strength distributions, so keep in mind that we are really comparing up transitions against down transitions.

The expression for charged current thermal strength detailed balance differs slightly from that for the neutral current channel.  We begin with the charged current thermal strength for transitions from nucleus $j$ to nucleus $k$, defined analogously to the neutral current.

\begin{equation}
\begin{split}
&S^{\pm}_{jk}\left(T,Q\right)=\\
&\frac{1}{G_j(T)}\int_0^\infty dE_j(2J_j+1)\rho_j(E_j,J_j)e^{-E_j/T}B^{\pm}_{jk}(E_j,Q)
\end{split}
\label{eq:cc_thermal_strength}
\end{equation}
The plus (minus) signs in the superscripts correspond to isospin-raising (-lowering) transitions, $G_j(T)$ is the partition function of nucleus $j$, and $Q$ is the total nuclear transition energy from an initial state in nucleus $j$ with energy $E_j$ to a final state in nucleus $k$ with energy $E_k$.

\begin{equation}
Q\equiv E_k+m_k-E_j-m_j
\label{eq:Q}
\end{equation}
We extend the observations of references \cite{thomas:1964,gg:1967,thomas:1968} to charged current interactions:

\begin{equation}
\begin{split}
&(2J_j+1)\rho_j(E_j,J_j)B^{\pm}_{jk}(E_j,Q)=\\
&(2J_k+1)\rho_k(E_k,J_k)B^{\mp}_{kj}(E_k,-Q).
\end{split}
\end{equation}
Substituting into equation \ref{eq:cc_thermal_strength} and using equation \ref{eq:Q} to substitute $E_k$ for $E_j$ gives

\begin{equation}
\begin{split}
&S^{\pm}_{jk}\left(T,Q\right)=\frac{1}{G_j(T)}\int_{Q-\Delta m}^\infty dE_k\\
&(2J_k+1)\rho_k(E_k,J_k)e^{-(E_k-Q+\Delta m)/T}B^{\mp}_{kj}(E_k,-Q)
\end{split}
\end{equation}
where $\Delta m\equiv m_k-m_j$.  The lower limit of this integral must obviously be at least zero (even if $Q-\Delta M<0$) since nuclear excitation can never be less than zero.  $E_k=Q-\Delta m$ corresponds to $E_j=0$, so any values of $E_k<Q-\Delta m$ correspond to values of $E_j<0$, which is unphysical.  If we interpret this as meaning that $B^{\mp}_{kj}(E_k<Q-\Delta m,-Q)=0$, then we may simply consider the lower limit on the integral to be zero.  This yields

\begin{equation}
\begin{split}
&S^{\pm}_{jk}\left(T,Q\right)=\frac{1}{G_j(T)}e^{(Q-\Delta m)/T}\\
&\times\int_{0}^\infty dE_k(2J_k+1)\rho_k(E_k,J_k)e^{-E_k/T}B^{\mp}_{kj}(E_k,-Q)\\
&=\frac{G_k(T)}{G_j(T)}e^{(Q-\Delta m)/T}S^{\mp}_{kj}(T,-Q).
\end{split}
\label{eq:cc_balance_asym}
\end{equation}
Rewriting to match the format of equation \ref{eq:nc_balance2}, we at last arrive at the expression for the detailed balance of thermal charged current strength.

\begin{equation}
\begin{split}
&e^{-\left(Q-\Delta m\right)/2T}G_j(T)S^{\pm}_{jk}\left(T,Q\right)=\\
&e^{\left(Q-\Delta m\right)/2T}G_k(T)S^{\mp}_{kj}\left(T,-Q\right)
\end{split}
\label{eq:cc_balance}
\end{equation}
Unlike the neutral current channel, this expression includes factors of each nuclear partition function because the initial and final nuclei are not identical.  

Following reference \cite{mf:2016}, we define the imbalance $I$ between two positive quantities $A$ and $B$ as

\begin{equation}
I(A,B) = \frac{A-B}{A+B}.
\label{eq:imbalance}
\end{equation}
We will use imbalance to compare quantities throughout this work, as it has some advantages over other traditional quantities of comparison, e.g., ratios and differences.  First, imbalance is (anti-)symmetric in $A$ and $B$.  Second, it is finite if either argument is zero.  Both of these qualities make it preferable to a ratio.  Third, it gives a measure of relative inequality, rather than absolute inequality, making it preferable to a difference when comparing quantities of arbitrary size such as transition strengths.  Fourth, it is bounded above and below, making it convenient to plot.

Imbalance has the disadvantage of not allowing an immediate, intuitive direct comparison (for example, ``$A$ is 4 times as large as $B$''), but this lack of intuition may simply be because we aren't yet used it.  It does have a mechanical analog: two objects with masses $m_1$ and $m_2$ suspended under gravity on either side of a physicist's pulley (massless, frictionless) will accelerate at $g\times I(m_1,m_2)$.  In any case, this particular difficulty with intuition is inconsequential when we are more concerned with trends than precise numbers.

\section{Shell Model Calculations}
\label{sec:calculations}

We used the shell model code OXBASH \cite{oxbash} and the USDB Hamiltonian \cite{br:2006} to compute nuclear energy levels and transition strengths for both charged current and neutral current transitions in several nuclei.  Where available, we used experimental energies and strengths.

The modification of the Brink-Axel hypothesis proposed in reference \cite{mfb:2014} is to include all states individually up to a cutoff energy and combine several states above the cutoff into a single high energy average state that carries the remainder of the thermal statistical weight.  This prescription is effective for computing electron capture and energy loss rates, but the lack of states at very high initial nuclear energy renders it unsatisfactory at producing neutrino energy spectra at very high neutrino energy.  We must include initial states with energies above the high energy average state to see the neutrino spectrum at very high neutrino energy, particularly in the neutral current channel where all of the neutrino energy comes from the nucleus (as opposed to the charged current channel, where incoming leptons provide some of the energy).

Therefore, we further modified the technique of reference \cite{mfb:2014} and used the following approach.  We considered all initial nuclear states individually up to 15 MeV.  We took as lower bin edges 15, 16, ..., 20 MeV.  In each of these bins, we computed transition strengths for the 20 lowest states.  We averaged the states in each bin together (weighting by spin degeneracy) to create an average state in that bin.  We assigned to each average state the total thermal statistical weight corresponding to states in its bin.

For the initial states under 15 MeV excitation (those considered individually), we computed the transition strengths to all final states in the daughter nucleus with total energy less than 35 MeV above the parent nucleus ground state.  For the binned initial states, we computed transition strengths to all final states below 20 MeV above the respective lower bin edge.  This likely clips the high transition energy tail of the highest energy initial states' strength distributions, but this is mitigated in three ways.  First, there is very little strength above 20 MeV transition energy.  Second, the high energy initial states are sparsely populated.  Third, transitions with large, positive transition energy require incoming particles with correspondingly high energy.  For charged current reactions, this can only happen at an appreciable rate late in stellar core collapse when the electron Fermi energy is extremely high, and for the neutral current, after core bounce, when energetic neutrinos are produced.

We converted the resulting strength distributions to thermal strengths according to equations \ref{eq:nc_thermal_strength} and \ref{eq:cc_thermal_strength} and compared the forward and reverse strengths by computing their imbalance (equation \ref{eq:imbalance}); recall that reverse reaction strength distributions are mirrored about $Q=0$ so that we can compare up transitions to down transitions.  We also computed the concomitant imbalance between the left and right hand sides of the detailed balance expressions (equations \ref{eq:nc_balance2} and \ref{eq:cc_balance}), which for the sake of brevity we term ``detailed imbalance''.

\begin{figure}
\includegraphics[scale=.5]{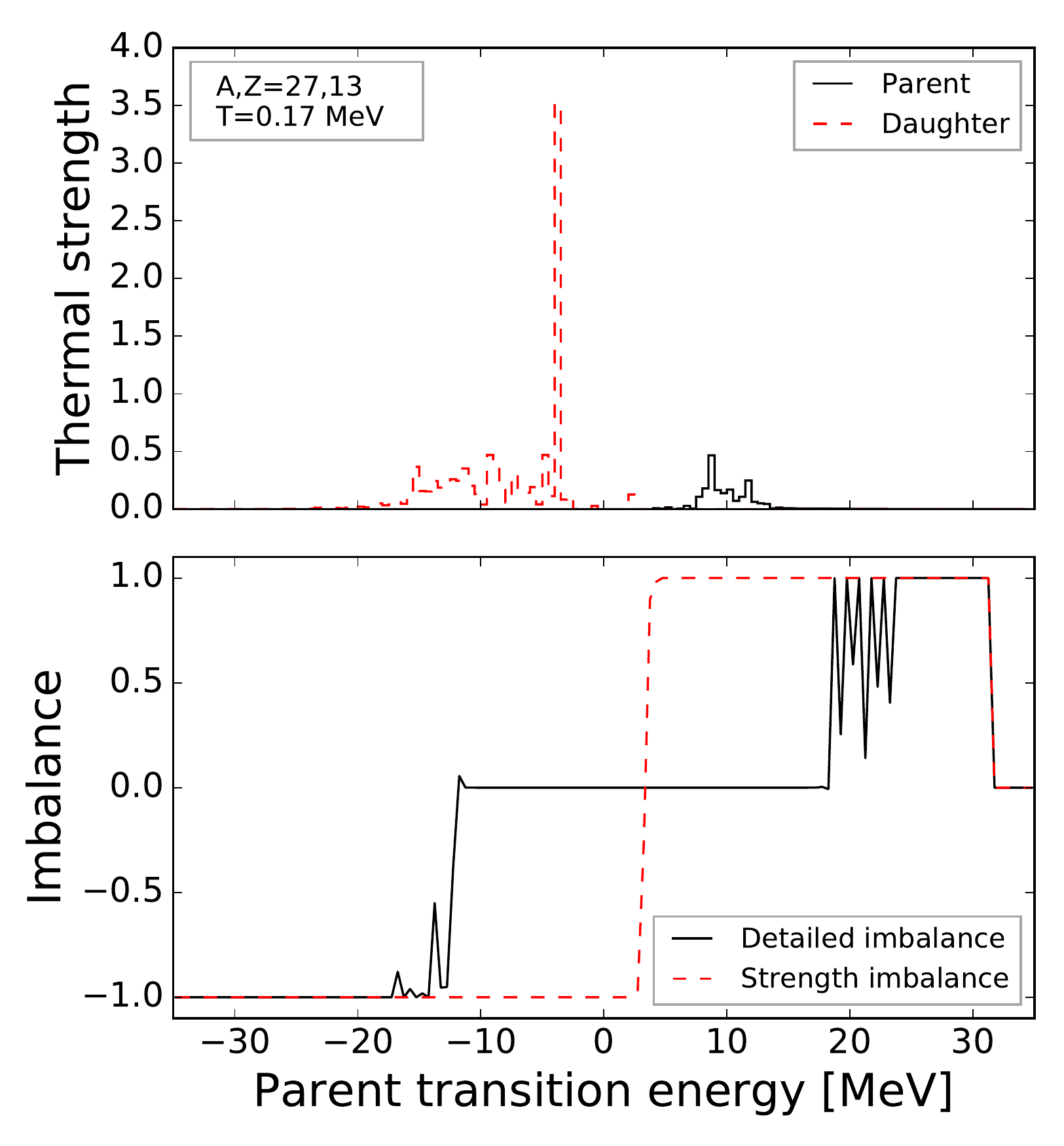}
\caption{$^{27}$Al isospin-lowering charged current strength, reverse reaction strength, and imbalance at temperature $T=0.17$ MeV.  The upper panel shows the thermal strengths for the forward (solid lines) and reverse (daughter [$^{27}$Mg] to parent, dashed lines) reactions, and the lower panel shows the imbalance between the left- and right-hand sides of equation \ref{eq:cc_balance} (``detailed imbalance'', solid lines) and in the thermal strength (dashed lines).  The large peak in the daughter thermal strength is from Fermi transitions.}
\label{fig:27al_imbalance_gt-_017}
\end{figure}

\begin{figure}
\includegraphics[scale=.5]{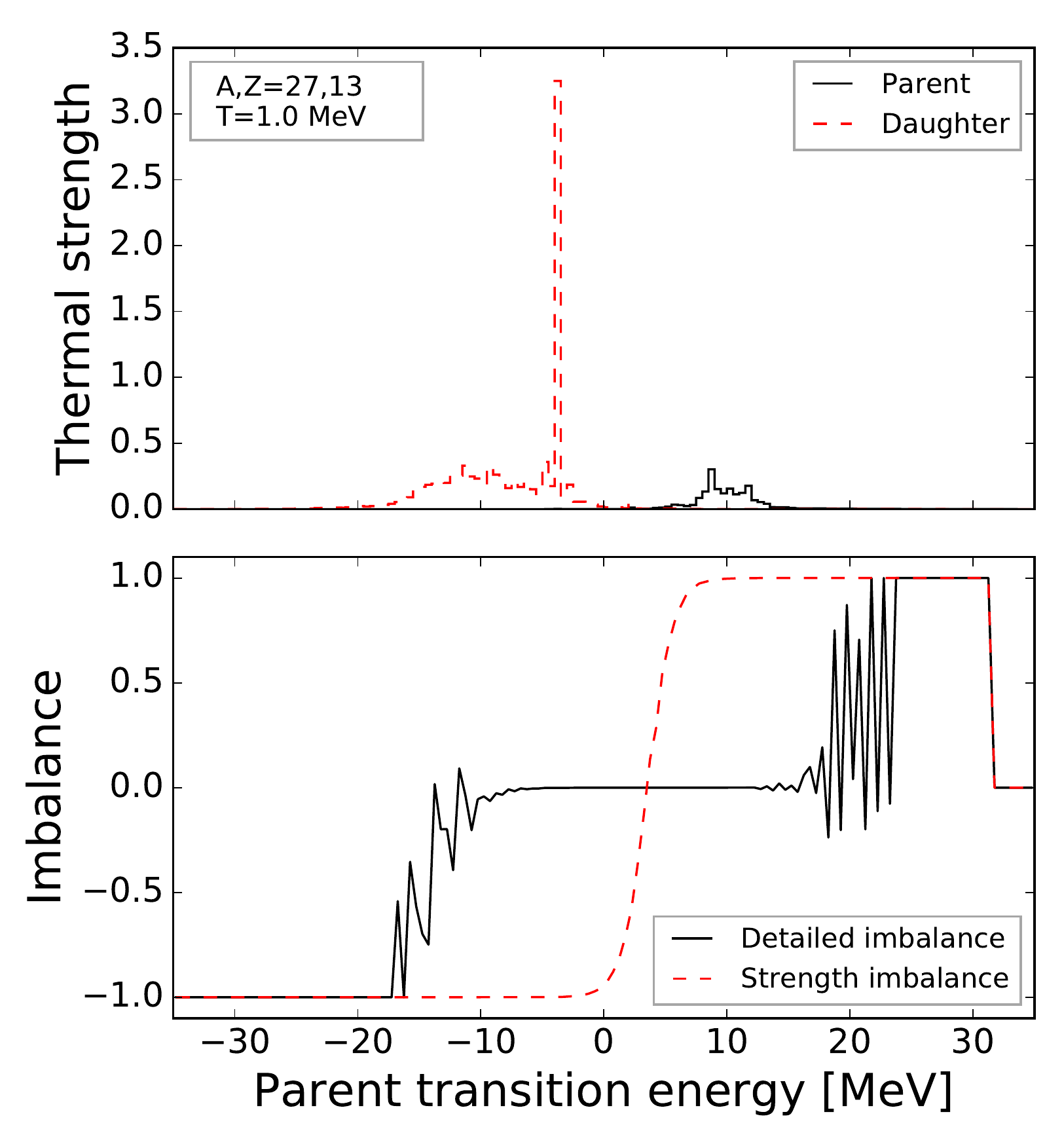}
\caption{$^{27}$Al isospin-lowering charged current strength and imbalance (as in figure \ref{fig:27al_imbalance_gt-_017}) at temperature $T=1.0$ MeV.}
\label{fig:27al_imbalance_gt-_100}
\end{figure}

Figures \ref{fig:27al_imbalance_gt-_017} and \ref{fig:27al_imbalance_gt-_100} show the thermal strengths for isospin-lowering charged current interactions on $^{27}$Al and the reverse reactions on $^{27}$Mg, the imbalance in the strengths, and the detailed imbalance at temperatures $T=0.17$ MeV and $T=1.0$ MeV, respectively.  The tails of the strength distributions fall off rapidly above $\sim$ 15 MeV transition energy (-15 MeV in the reverse reaction), justifying the truncation in transition energy described above.

At both temperatures, the thermal strength imbalance is overwhelming more than a few MeV from 0 transition energy.  Most importantly, the detailed imbalance is near zero within $\sim 10$ MeV of the region where the thermal strength imbalance is not extreme.  Therefore, we conclude that detailed imbalance only occurs far out on the tail of one or the other of the thermal strength distributions where there is very little strength, which is to say, in regions that do not contribute much to the overall rate of that interaction.

\begin{figure}
\includegraphics[scale=.5]{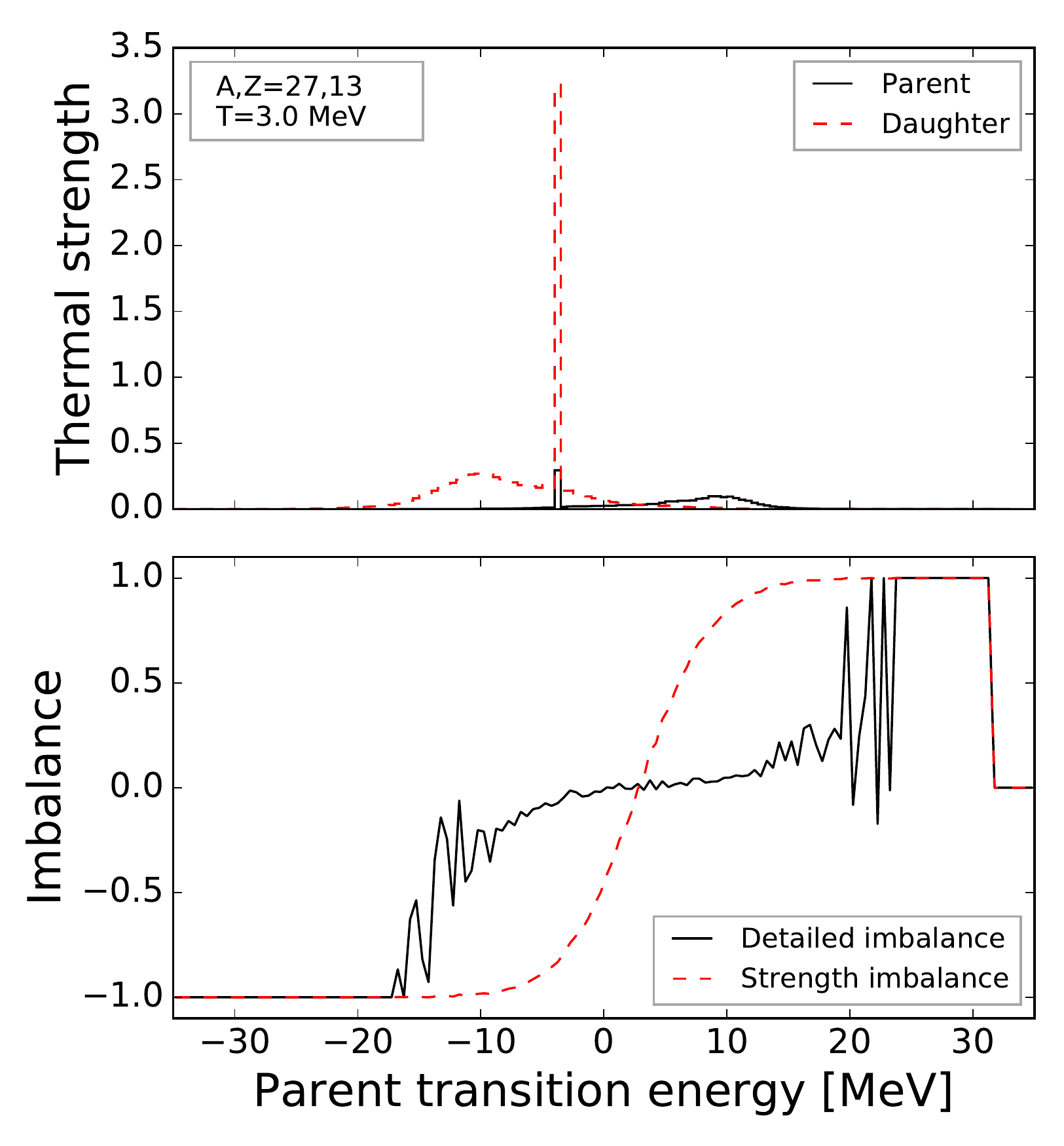}
\caption{$^{27}$Al isospin-lowering charged current strength and imbalance (as in figure \ref{fig:27al_imbalance_gt-_017}) at temperature $T=3.0$ MeV.  Even at this extreme temperature, detailed imbalance is only large where thermal strength imbalance is very large, though the gap between large detailed imbalance and small thermal strength imbalance has become relatively narrow.}
\label{fig:27al_imbalance_300}
\end{figure}

Figure \ref{fig:27al_imbalance_300} shows the same distributions as figure \ref{fig:27al_imbalance_gt-_017}, but at the extreme temperature of $T=3.0$ MeV.  As higher energy nuclear states become populated, the region where thermal strength imbalance is small broadens.  There remains a gap of a few MeV between this region and the region in which detailed imbalance is large, but the gap is somewhat tenuous, and we might expect some amount of disagreement in the reaction rates as a consequence; we will touch on this later.

\begin{figure}
\includegraphics[scale=.5]{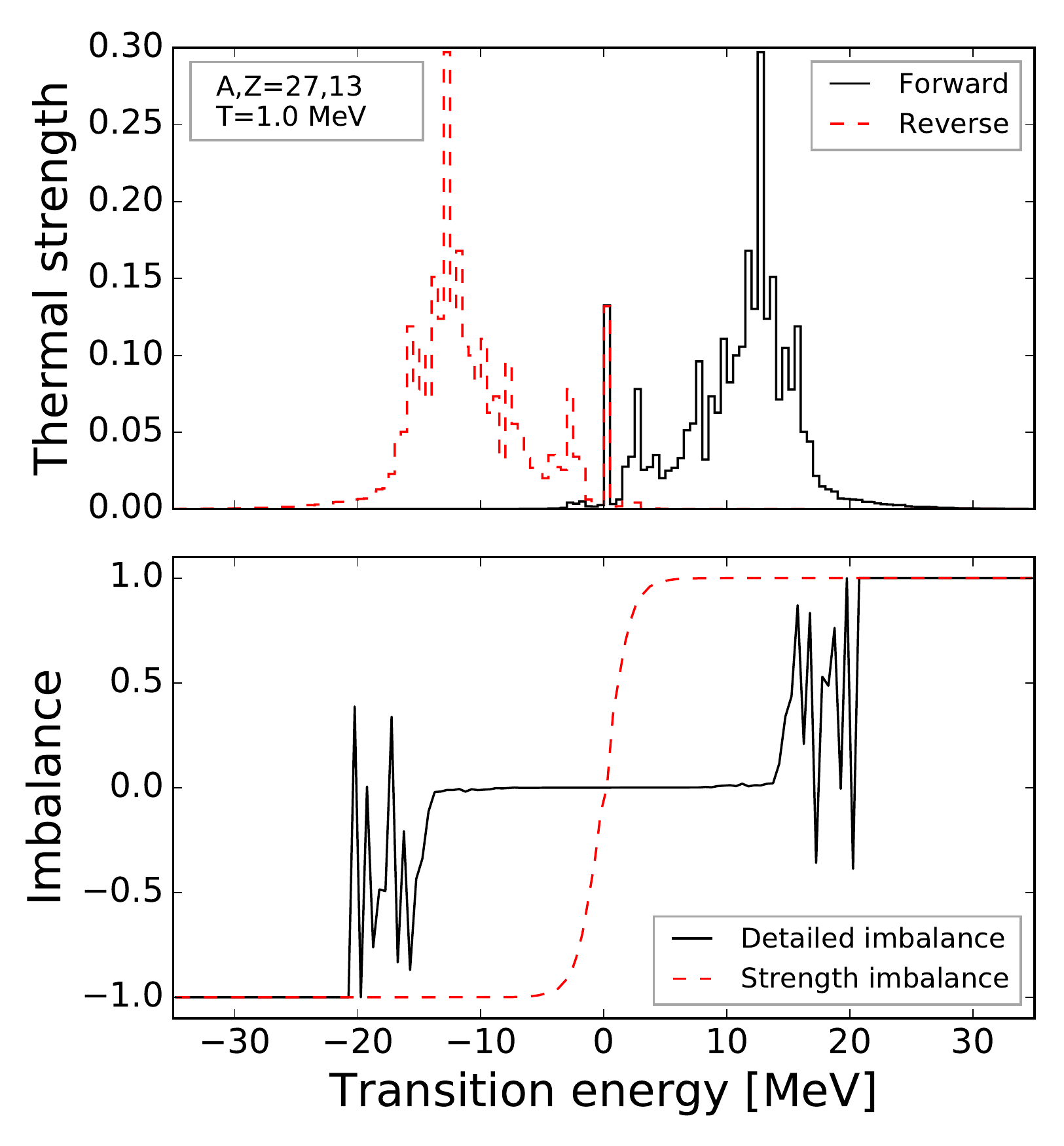}
\caption{$^{27}$Al neutral current strength and imbalance at temperature $T=1.0$ MeV.  All quantities are defined the same as in figure \ref{fig:27al_imbalance_gt-_017}, though here we use ``Forward'' and ``Reverse'' since the parent and daughter nuclei are identical.}
\label{fig:27al_imbalance_gt3}
\end{figure}

Figure \ref{fig:27al_imbalance_gt3} shows the forward and reverse neutral current thermal strength distributions and imbalances for $^{27}$Al at temperature $T=1.0$ MeV.  We draw the same conclusions as for the charged current channel: the detailed imbalance is only large far from where the thermal strength imbalance is not also extremely large.

\begin{figure}
\includegraphics[scale=.5]{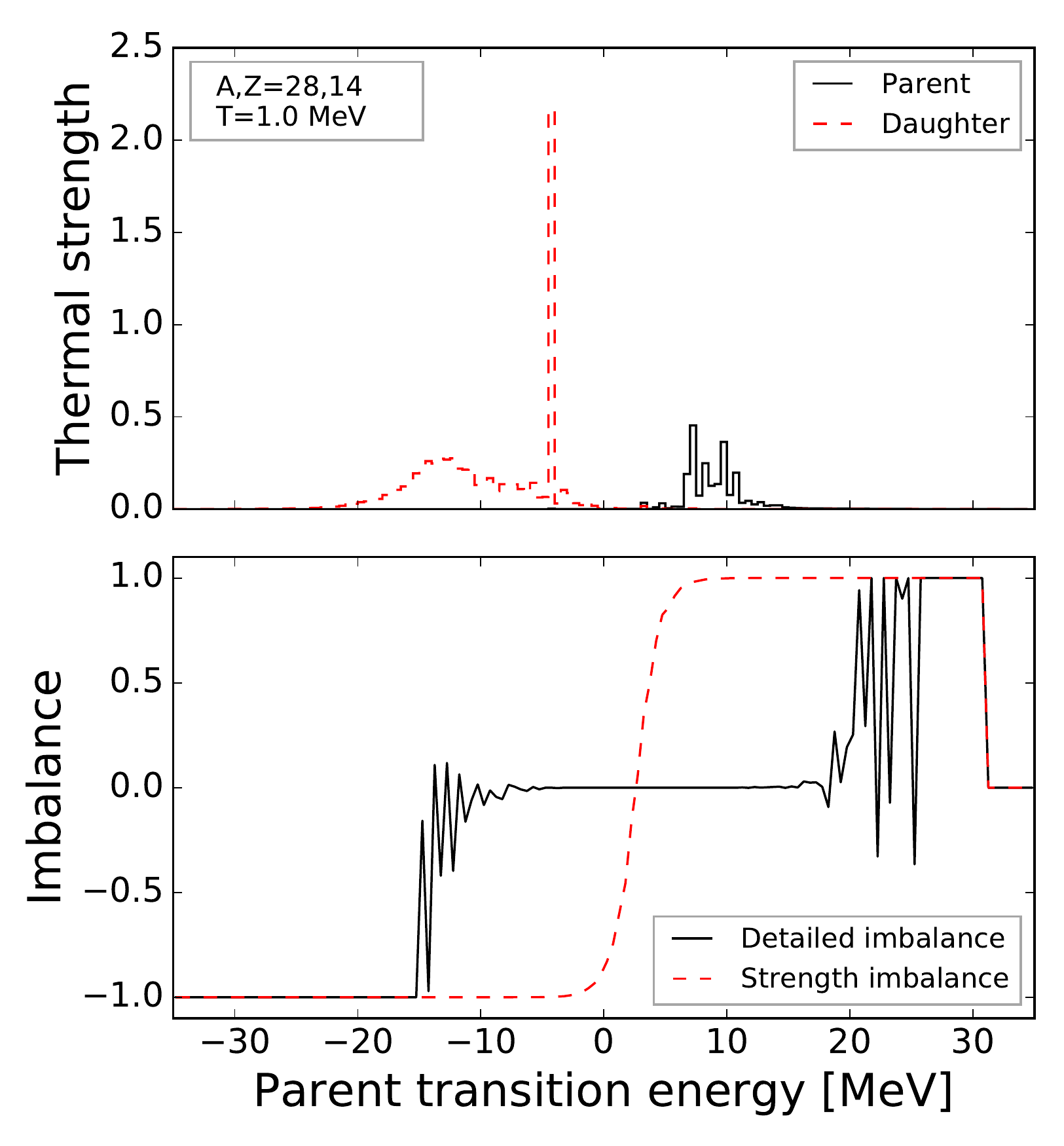}
\caption{$^{28}$Si isospin-lowering charged current strength and imbalance (as in figure \ref{fig:27al_imbalance_gt-_017}) at temperature $T=1.0$ MeV.  In this case, the daughter nucleus is $^{28}$Al.}
\label{fig:28si_imbalance_gt-}
\end{figure}

\begin{figure}
\includegraphics[scale=.5]{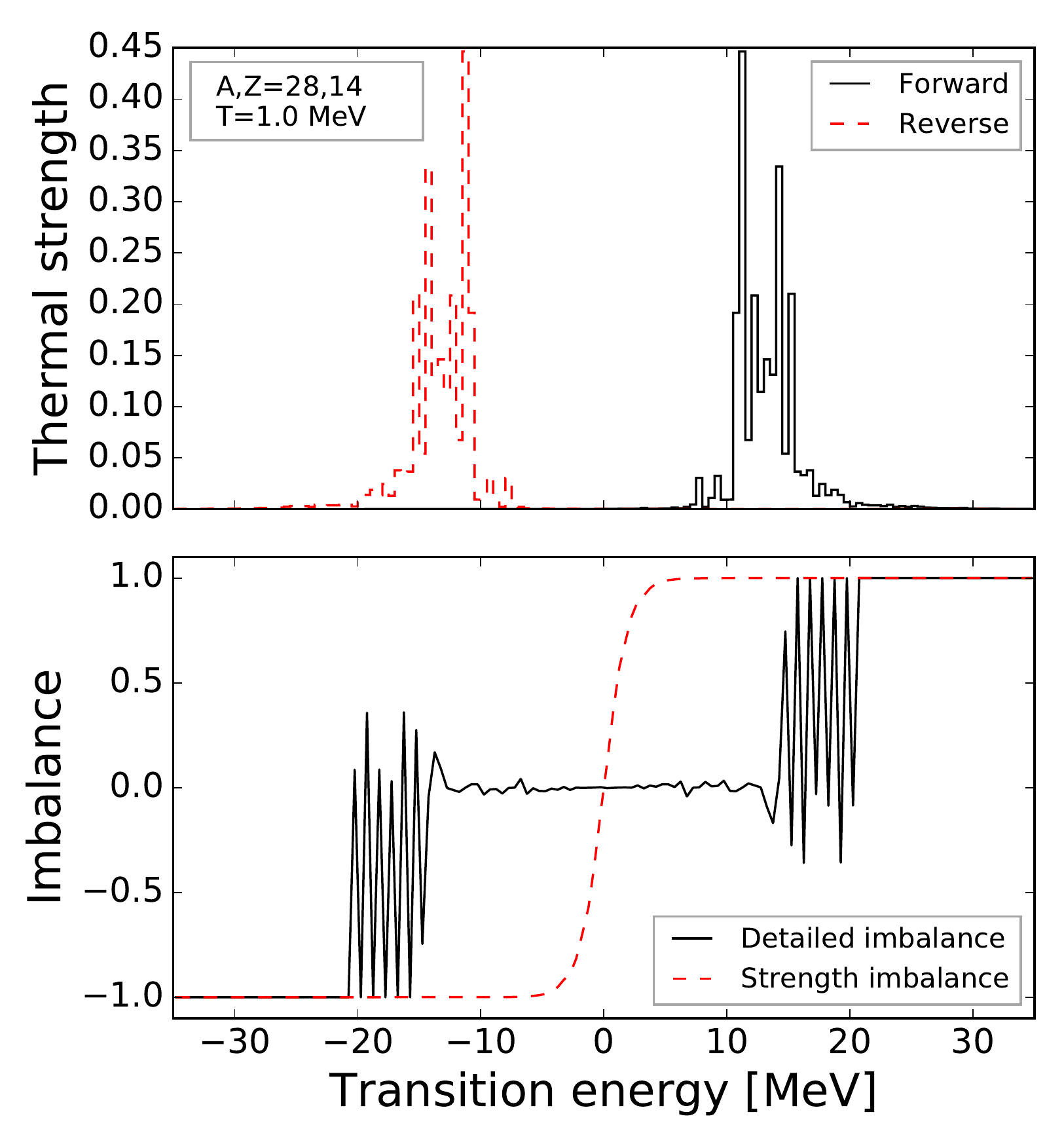}
\caption{$^{28}$Si neutral current strength and imbalance (as in figure \ref{fig:27al_imbalance_gt3}) at temperature $T=1.0$ MeV.}
\label{fig:28si_imbalance_gt3}
\end{figure}

\begin{figure}
\includegraphics[scale=.5]{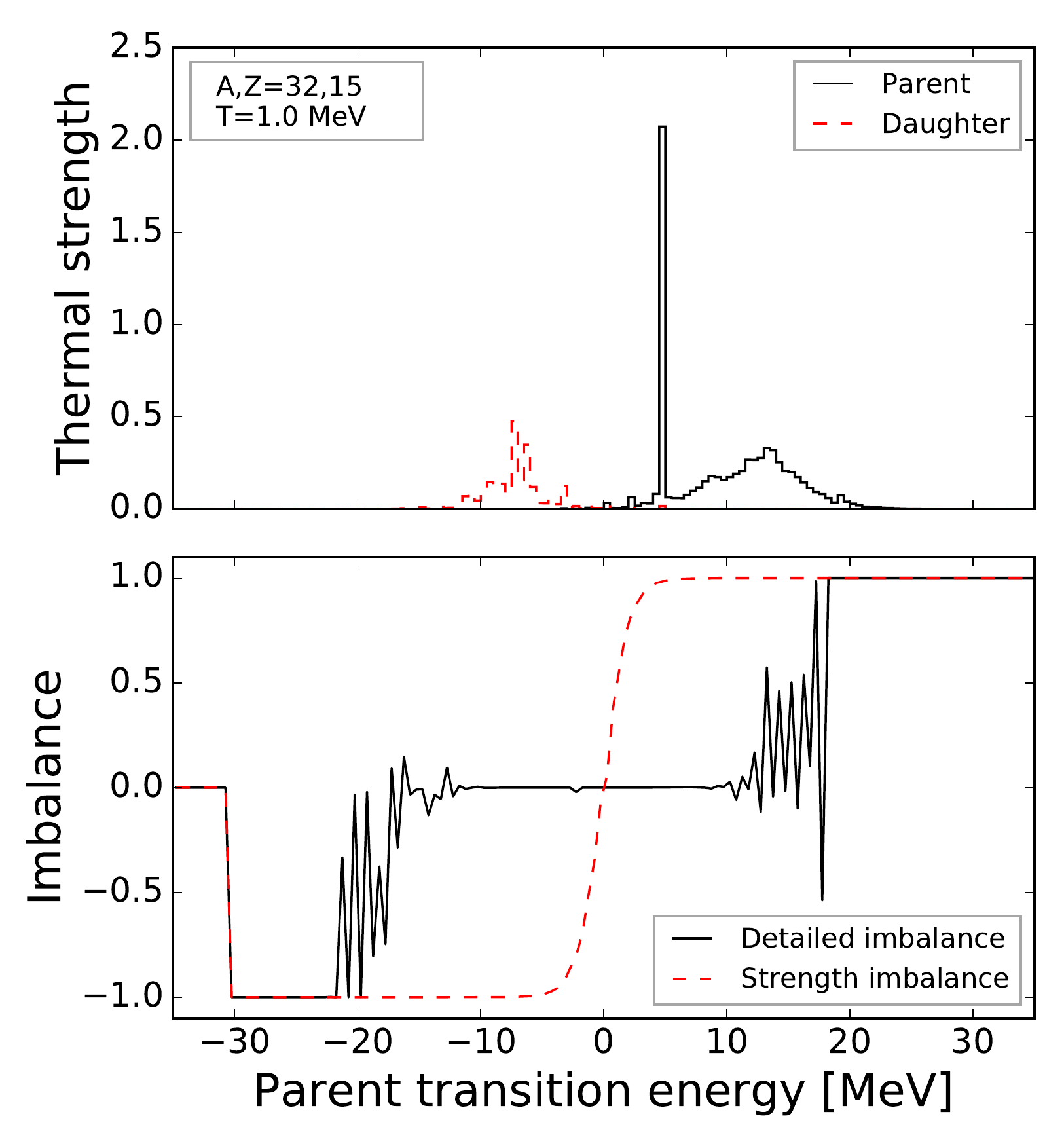}
\caption{$^{32}$P isospin-raising charged current strength and imbalance (as in figure \ref{fig:27al_imbalance_gt-_017}) at temperature $T=1.0$ MeV.  In this case, the daughter nucleus is $^{32}$S.}
\label{fig:32p_imbalance_gt+}
\end{figure}

\begin{figure}
\includegraphics[scale=.5]{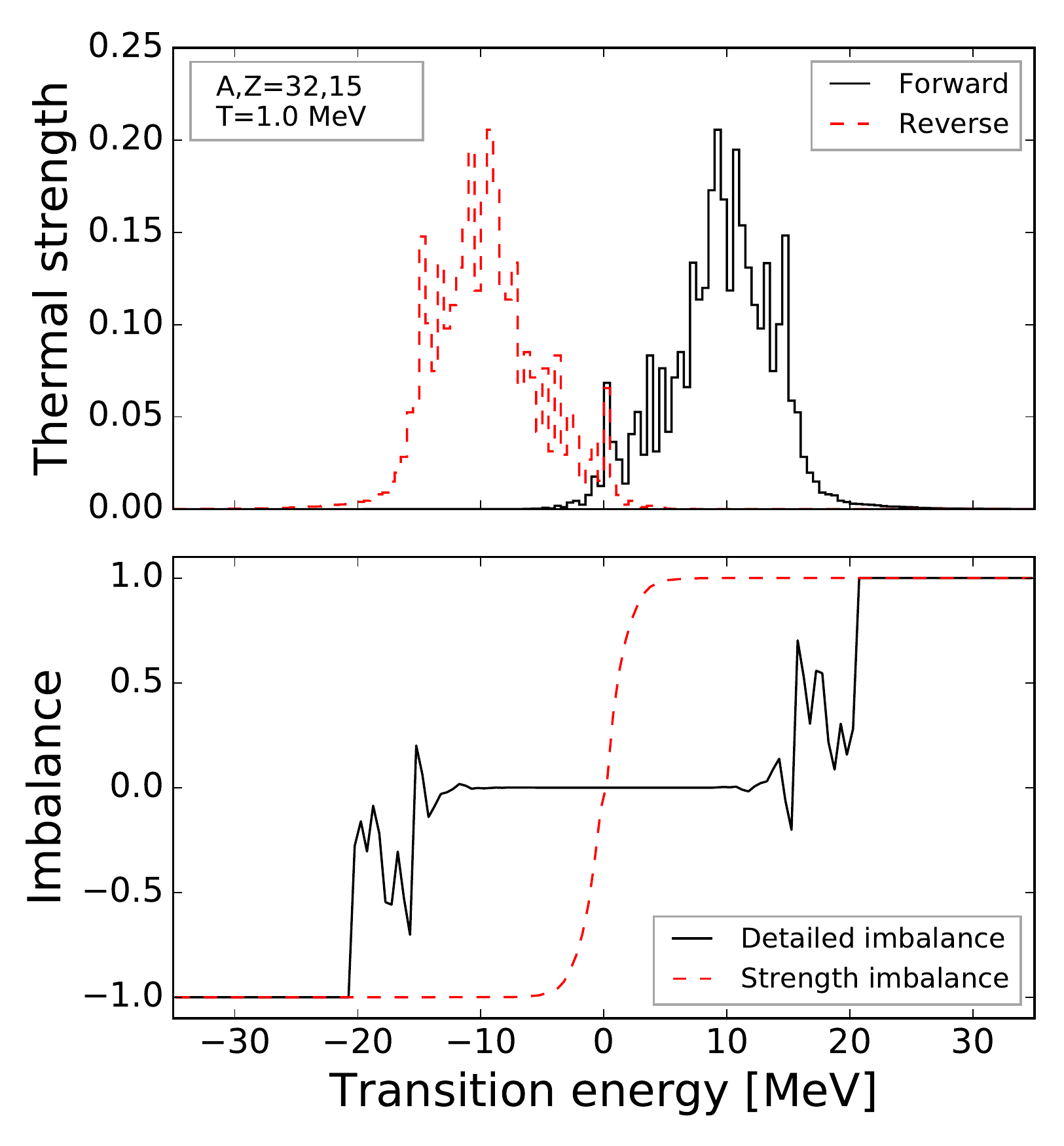}
\caption{$^{32}$P neutral current strength and imbalance (as in figure \ref{fig:27al_imbalance_gt3}) at temperature $T=1.0$ MeV.}
\label{fig:32p_imbalance_gt3}
\end{figure}

$^{27}$Al is an odd-even nucleus (odd number of protons, even number of neutrons); we wish to ascertain whether these results hold for other kinds of nuclei.  Figures \ref{fig:28si_imbalance_gt-} and \ref{fig:28si_imbalance_gt3} show the thermal strength and imbalance for the even-even nucleus $^{28}$Si at temperature $T=1.0$ MeV (charged current and neutral current, respectively), and figures \ref{fig:32p_imbalance_gt+} and \ref{fig:32p_imbalance_gt3} show the same for the odd-odd nucleus $^{32}$P.  These figures agree with the results for $^{27}$Al.

\section{Phase Space Factors}
\label{sec:phase_space}

Weak reaction rates $\lambda_{ij}$ between two nuclear states $\vert i\rangle$ and $\vert j\rangle$ can be expressed as the product of a base rate $\lambda_0$ (which contains physical constants), the transition strength $B_{ij}$ (which is a property of the nucleus), and a phase-space integral $f_{ij}$ (which accounts for the dynamics of incoming and outgoing particles).  Within a given channel, $\lambda_0$ and $B_{ij}$ will be the same for all reactions (though $B_{ij}$ will of course vary depending on the initial and final states), but the phase space integral can be qualitatively different for different reactions.  For example, in the charged current (CC) channel, electron capture (ec) and $\beta^-$ decay have identical relevant physical constants and matrix elements (up to a factor of $\frac{(2J_i+1)}{(2J_j+1)}$), but the phase space integral of the former must consider outgoing electron and neutrino blocking, while the latter must account for electron availability and outgoing neutrino blocking; equations \ref{eq:phase_space_ec} and \ref{eq:phase_space_beta} illustrate this.

\begin{equation}
\begin{aligned}
\lambda_{ij}^{ec}=&\lambda_0^{CC}B_{ij}^{-}f_{ij}^{ec} \\
f_{ij}^{ec}=&\int_{w_0}^\infty dw_e \int_{0}^\infty dw_\nu w_e^2 w_\nu^2 G(Z,w_e) \\
&\times\delta(w_e-w_\nu-q_{ij}) \\
&\times f_e(1-f_\nu)
\end{aligned}
\label{eq:phase_space_ec}
\end{equation}

\begin{equation}
\begin{aligned}
\lambda_{ij}^{\beta^-}=&\lambda_0^{CC}B_{ij}^{+}f_{ij}^{\beta^-}=\lambda_0^{CC}\frac{2J_j+1}{2J_i+1} B_{ji}^{-}f_{ij}^{\beta^-} \\
f_{ij}^{\beta^-}=&\int_{1}^{q_{ij}}dw_e\int_{0}^{q_{ij}-1}dw_\nu w_e^2w_\nu^2G(Z,w_e) \\
&\times\delta(w_e+w_\nu+q_{ij}) \\
&\times (1-f_e)(1-f_\nu)
\end{aligned}
\label{eq:phase_space_beta}
\end{equation}

All energies are in units of electron mass $m_e$.  $B_{ij}^{-}$ ($B_{ij}^{+}$) is the transition strength for isospin-lowering (raising) transitions, $w$ is the electron energy (including rest mass), $q$ is the transition energy (as defined in section \ref{sec:strength_imbalance}, but expressed in units of $m_e$), $G(Z,w)$ is the Coulomb correction factor defined in reference \cite{ffn:1980}, and $f_e$ and $f_\nu$ are the electron and neutrino distribution functions, respectively.  The Dirac $\delta$ function ensures conservation of energy.  The lower limit $w_0$ in the electron capture integral is the greater of 1 and $q$ (since the incoming electron has at least its rest mass and must provide enough energy to the nucleus to make the transition).

In the neutral current ($NC$) channel, let us consider neutral current deexcitation ($de$) and its reverse reaction, neutrino pair absorption ($pa$).  These reactions proceed via the isospin z-projection of the Gamow-Teller interaction ($GT_3$).  Equations \ref{eq:phase_space_de} and \ref{eq:phase_space_pa} show these rates and phase space integrals.

\begin{equation}
\begin{aligned}
\lambda_{ij}^{de}=&\lambda_0^{NC}B_{ij}^{GT_3}f_{ij}^{de} \\
f_{ij}^{de}=&\int_0^q\int_0^qw_\nu^2w_{\overline{\nu}}^2\delta(w_\nu+w_{\overline{\nu}}-q) \\
&\times (1-f_\nu)(1-f_{\overline{\nu}})dw_\nu dw_{\overline{\nu}}
\end{aligned}
\label{eq:phase_space_de}
\end{equation}

\begin{equation}
\begin{aligned}
\lambda_{ij}^{pa}=&\lambda_0^{NC}B_{ij}^{GT_3}f_{ij}^{pa} \\
f_{ij}^{pa}=&\int_0^q\int_0^qw_\nu^2w_{\overline{\nu}}^2\delta(w_\nu+w_{\overline{\nu}}-q) \\
&\times f_\nu f_{\overline{\nu}}dw_\nu dw_{\overline{\nu}}
\end{aligned}
\label{eq:phase_space_pa}
\end{equation}

As in equations \ref{eq:phase_space_ec} and \ref{eq:phase_space_beta}, all energies are in units of $m_e$.  Here $w_\nu$ and $w_{\overline{\nu}}$ are the neutrino and antineutrino energies, respectively, while $f_\nu$ and $f_{\overline{\nu}}$ are the respective distribution functions.

As we will see below, the qualitative differences in the phase space factors between forward and reverse reactions has a profound effect on reaction rates when the core is out of weak equilibrium.

\section{Reaction Rates}
\label{sec:rates}

Now that we have a handle on the thermal transition strengths and the phase space factors, we wish to investigate the effects of thermal strength detailed imbalance on reaction rate calculations.  Table \ref{table:rates} shows the isospin-raising reaction rates for $^{32}$P over a wide range of temperatures and densities; the specific values of temperature and density were selected for easy comparison with Oda et al \cite{oda-etal:1994}.  The temperature is listed in units of $10^9$ Kelvin ($T_9$), and the density is listed as the product of of the mass density $\rho$ in g/cm$^3$ and the electron fraction (electrons per baryon) $Y_e$.  Two rates are listed for each entry (temperature, density, and reaction): the upper value is computed from the $^{32}$P isospin-raising strength found using the technique described in section \ref{sec:calculations}, and the lower is computed with the same technique, but using the $^{32}$S isospin-lowering strength and applying equation \ref{eq:cc_balance_asym}; the second method obeys detailed balance between $^{32}$P and $^{32}$S thermal strengths by construction.

\begin{table}
\begin{tabular}{c | c | c | c | c}
$T_9$ & Reaction & $\rho Y_e=10$ & $10^5$ & $10^{10}$ \\
\hline
\multirow{2}{*}{0.1} & e$^+$ cap & -58.060 & -63.576 & -999 \\
& & -58.060 & -63.575 & -999 \\
& e$^-$ dec & -7.751 & -7.759 & -999 \\
& & -7.751 & -7.759 & -999 \\
\hline
\multirow{2}{*}{0.7} & e$^+$ cap & -11.404 & -14.838 & -91.388 \\
& & -11.404 & -14.838 & -91.387 \\
& e$^-$ dec & -7.914 & -7.920 & -70.049 \\
& & -7.914 & -7.920 & -70.049 \\
\hline
\multirow{2}{*}{3.0} & e$^+$ cap & -6.814 & -6.873 & -25.452 \\
& & -6.814 & -6.873 & -25.452 \\
& e$^-$ dec & -5.912 & -5.912 & -18.796 \\
& & -5.912 & -5.912 & -18.796 \\
\hline
\multirow{2}{*}{10.0} & e$^+$ cap & -3.694 & -3.695 & -9.176 \\
& & -3.695 & -3.696 & -9.177 \\
& e$^-$ dec & -3.951 & -3.951 & -7.035 \\
& & -3.951 & -3.951 & -7.035 \\
\hline
\multirow{2}{*}{30.0} & e$^+$ cap & -0.328 & -0.328 & -1.859 \\
& & -0.570 & -0.570 & -2.099 \\
& e$^-$ dec & -2.128 & -2.128 & -2.859 \\
& & -2.154 & -2.154 & -2.865 
\end{tabular}
\caption{Isospin-raising reaction rates for $^{32}$P.  For each entry (temperature, $\rho Y_e$, and reaction), the upper rate is computed using transition strengths found from the technique described in section \ref{sec:calculations}, and the lower is calculated with that technique, but using the $^{32}$S isospin-lowering strength and applying equation \ref{eq:cc_balance_asym}.  The rates are listed as log(rate), where the rate is s$^{-1}$ baryon$^{-1}$.  The temperature is in units of $T_9$ ($10^9$ Kelvin) and $\rho Y_e$ is in g/cm$^2$.}
\label{table:rates}
\end{table}

Up to temperature $T_9=10$ (0.862 MeV), both methods agree to high precision.  At $T_9=30$ (2.585 MeV), however, the detailed balance method predicts $e^+$ capture rates that are lower than the direct calculation method by a factor of $\sim 1.74$ at all densities.  In fact, this implies that in this case, explicitly imposing detailed balance causes us to miss some strength at high temperature!

\begin{figure}
\includegraphics[scale=.5]{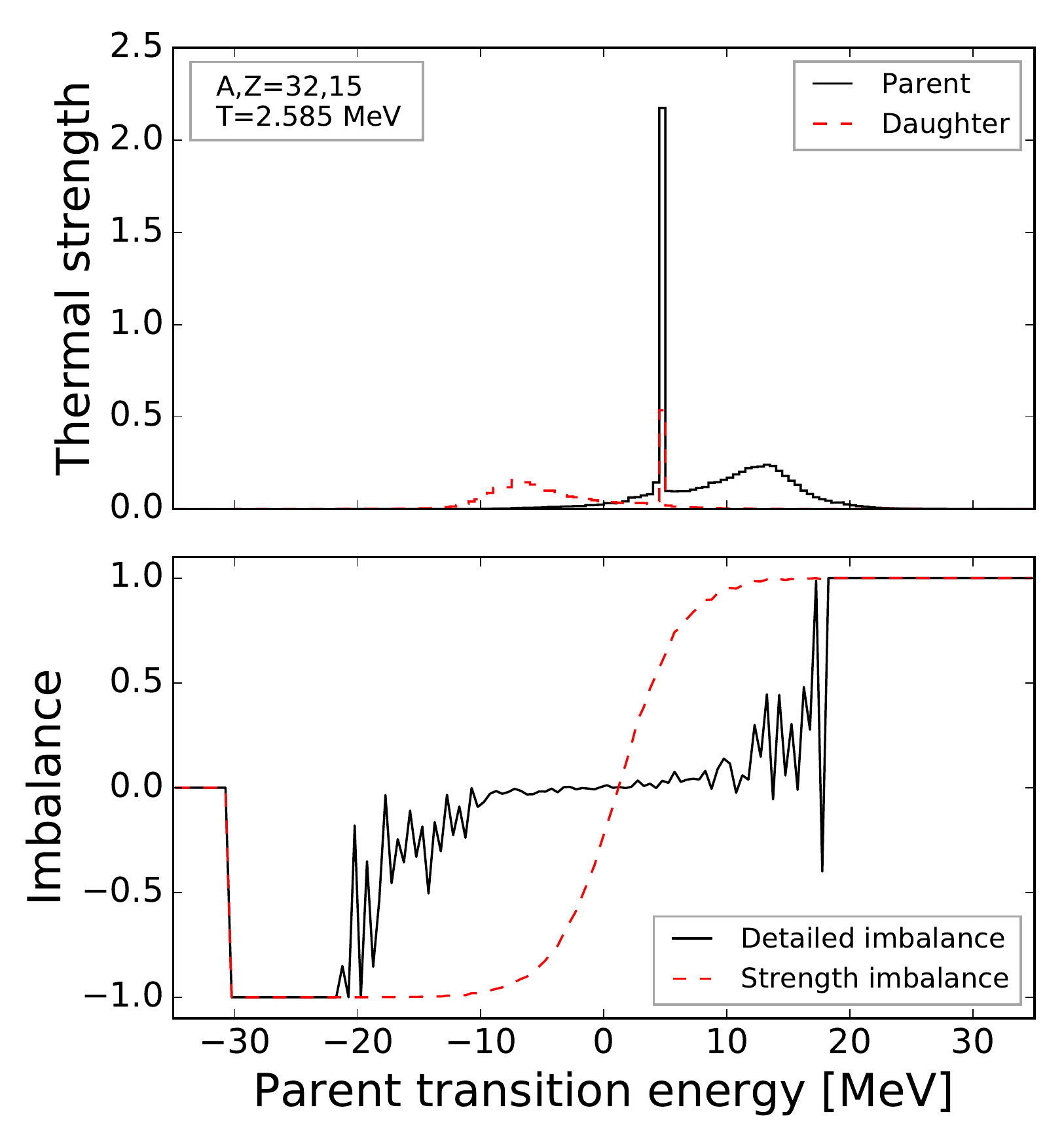}
\caption{$^{32}$P isospin-raising charged current strength and imbalance (as in figure \ref{fig:27al_imbalance_gt-_017}) at temperature $T_9=30$ (2.585 MeV).  As in figure \ref{fig:27al_imbalance_300}, the region of small strength imbalance is broad.}
\label{fig:32p_imbalance_gt+_259}
\end{figure}

Figure \ref{fig:32p_imbalance_gt+_259} shows the thermal isospin-raising strength and imbalance for $^{32}$P at $T_9=30$.  These curves are qualitatively similar to those in figure \ref{fig:27al_imbalance_300}, which was also at an extreme temperature.

\begin{figure}
\includegraphics[scale=.5]{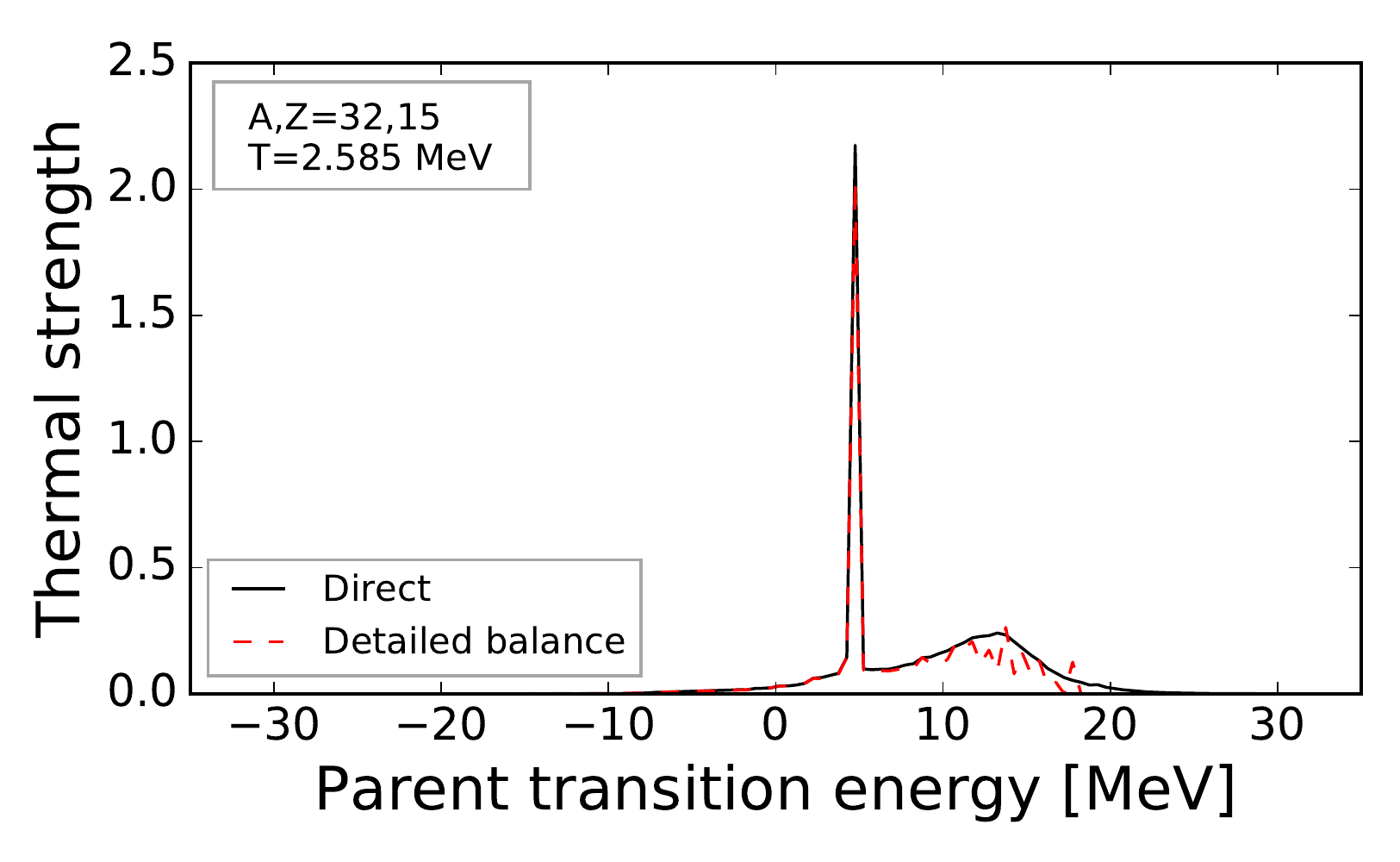}
\caption{$^{32}$P isospin-raising thermal strength as computed using the left- (``Direct'') and right-hand (``Detailed balance'') sides of equation \ref{eq:cc_balance_asym}.  The detailed balance method misses some strength at high transition energy, leading to an underestimate of the positron capture rate.}
\label{fig:32p_strength_gt+_259}
\end{figure}

Figure \ref{fig:32p_strength_gt+_259} shows the left- and right-hand sides of equation \ref{eq:cc_balance_asym}; the right-hand side is the strength used to compute the lower values in table \ref{table:rates}.  The missing high temperature strength is clear to see in the broad peak from $\sim 10$--$20$ MeV transition energy.  Evidently, at extremely high temperatures, the positrons have a long enough high-energy tail that the strength in this energy region contributes significantly, leading to an underestimate of the rate by the detailed balance method.  

To further understand this discrepancy, consider that the direct method computes the strength computes strength \emph{from parent states}, while the detailed balance method computes strength \emph{to daughter states}.  High-energy parent states are suppressed by the Boltzmann factor, but high-energy daughter states are always reachable as long as the incoming lepton has enough energy.  The consequence is that the approximation we use to directly compute thermal strengths will be very good to high temperatures, but the relatively sparse sampling of high-energy states can lead to errors when using the detailed balance method.  We thus conclude that the method of section \ref{sec:calculations} adequately satisfies detailed balance of charged current reaction strengths for temperatures below 1 MeV, and that even up to very high temperatures, the violation of detailed balance does not lead to extreme disagreement in the computed rates.  Furthermore, imposing detailed balance may in fact lead to some missing strength.

\section{Discussion}
\label{sec:discussion}

We must keep in mind the distinction between detailed balance (which refers to the relationship between forward and reverse transition strengths) and weak equilibrium (which refers to equality between forward and reverse reaction rates).  While a realistic model must obey detailed balance of strength, violations thereof may be unimportant if the system under consideration is far from equilibrium, since under such conditions, reactions proceed much faster in one direction than the other.

Consider, for example, equations \ref{eq:phase_space_ec} and \ref{eq:phase_space_beta}.  The blocking factor $1-f_e$ from electrons inhibits $\beta$ decay, while a high electron Fermi energy can greatly enhance electron capture.  As the core collapses, the increasing density pushes the electron Fermi energy from a few MeV to a few tens of MeV, driving electron capture and blocking $\beta$ decay, and all the while pushing the core material far from $\beta$ equilibrium (until neutrinos become trapped, allowing neutrino captures to occur at a thermal equilibrium rate).  In this situation, where electron capture overwhelms $\beta$ decay, we must be confident of electron capture rates, but whether the forward and reverse strengths closely obey detailed balance is not particularly helpful in understanding the dynamics of the collapse.

Now consider equations \ref{eq:phase_space_de} and \ref{eq:phase_space_pa}.  The rate formula for deexcitation into neutrino pairs includes blocking factors for the outgoing neutrino and antineutrino, while the reverse reaction--pair absorption--requires an incoming neutrino-antineutrino pair.  Pre-collapse and until the neutrinos become trapped late in collapse, they stream freely out of the core, so no equilibrium population builds up, and $f_\nu=f_{\overline{\nu}}\sim 0$.  This means that pair emission proceeds uninhibited, but pair absorption essentially can't happen at all, and the relative strengths of the reactions are irrelevant.

We conclude that when computing rates of nuclear weak interactions, it is sufficient to directly compute the forward reaction to an appropriate precision without concerning ourselves about whether the particular method strictly obeys detailed balance (though we have already shown that the approach used here largely does).

\section{Acknowledgments}
\label{sec:acknowledgments}

I gratefully thank George M. Fuller, Yang Sun, and Surja K. Ghorui for fruitful discussions.  I also owe gratitude to Projjwal Banerjee, Alice Shih, and Joe Semmelrock for their input in writing this manuscript.  This research at Shanghai Jiao Tong University is supported by the National Natural Science Foundation of China (No. 11575112), by the National Key Program for S\&T Research and Development (No. 2016YFA0400501), and by the 973 Program of China (No. 2013CB834401).

\bibliography{../references}

\end{document}